# IntrinsicTimescales.jl: A Julia package to estimate intrinsic (neural) timescales (INTs) from time-series data

Yasir Çatal[1, 2], Georg Northoff[1, 2]

[1]Mind, Brain Imaging and Neuroethics Research Unit, University of Ottawa, Ontario, ON, Canada.

[2]University of Ottawa Institute of Mental Health Research, Ottawa, ON, Canada.

## Abstract

IntrinsicTimescales.jl is a Julia package to perform estimation of intrinsic neural timescales (INTs). INTs are defined as the time window in which prior information from an ongoing stimulus can affect the processing of newly arriving information[1–4]. INTs are estimated either from the autocorrelation function (ACF) or the power spectral density (PSD) of time-series data[5,6]. In addition to the model-free estimates of INTs, IntrinsicTimescales.jl offers implementations of novel techniques of timescale estimation via performing parameter estimation of an Ornstein-Uhlenbeck process with adaptive approximate Bayesian computation (aABC)[7,8] and automatic differentiation variational inference (ADVI)[9].

## Statement of Need

Intrinsic neural timescales (INTs) were found to be an important metric to probe the brain dynamics and function. On the neuroscientific side, INTs were found to follow the large-scale gradients in the cortex ranging from uni to transmodal areas including local and long-range excitation[10,11] and proxies of myelination[11,12]. From a cognitive science perspective, INTs were found to be related to reward[10], behavior[13,14], self[15], consciousness[16] among others. Proper estimation of INTs to make sure the estimates are not affected by limited data, missingness of the data and oscillatory artifacts is crucial. While several methods exist for estimating INTs, there is a lack of standardized, open-source tools that implement both traditional model-free approaches and modern Bayesian estimation techniques. Existing software solutions are often limited to specific estimation methods, lack proper uncertainty quantification, or are not optimized for large-scale neuroimaging data.

IntrinsicTimescales.jl addresses these limitations by providing a comprehensive, high-performance toolbox for INT estimation. The package implements both established model-free methods and novel Bayesian approaches, allowing researchers to compare and validate results across different methodologies with a simple API. Its implementation in Julia ensures computational efficiency, crucial for analyzing large neuroimaging datasets. The package's modular design facilitates easy extension and integration with existing neuroimaging workflows, while its rigorous testing and documentation make it accessible to researchers across different levels of programming expertise.

## Major Features

IntrinsicTimescales.jl provides the following features:

- Model-free methods: IntrinsicTimescales.jl offers a unified API for the following INT estimation methods, making it easy for the user to compare different estimation methods and allow flexibility in scripting.
    - ACW-50: Time to reach 0.5 in the ACF[5]
    - ACW-0: Time to reach 0.0 in the ACF[15,17]
    - ACW-e: Time to each 1/e in the ACF[18]
    - AUC: Area under the curve of the ACF from lag-zero to the lag where ACF reaches 0[19–22]
    - tau: The inverse decay rate of an exponential decay function fitted to the ACF[10,14]
    - knee: Estimation of the inverse decay rate of exponential decay function by fitting a Lorentzian to the PSD. By Wiener-Khinchine theorem, the decay rate of an exponential decay function is proportional to the knee frequency of the PSD[6,19]. Following the FOOOF package[6], we follow a 3-step procedure for estimating the knee frequency. First, a Lorentzian is fitted to the PSD. Secondly, the Lorentzian fit is subtracted and gaussians for oscillatory peaks are fitted to the residual PSD. Finally the gaussians are subtracted from base PSD and a Lorentzian is fitted one more time.
- Bayesian Parameter Estimation: These methods estimate the timescale as a parameter of a generative model involving an Ornstein-Uhlenbeck process. Generative models support missing data and oscillatory artifacts.
    - Adaptive Approximate Bayesian Computation (aABC) with population Monte Carlo[7,8]
- Automatic Differentiation Variational Inference (ADVI) through Turing.jl[9]
- Specialized ACF and PSD calculations:
    - In the case of no missing data, IntrinsicTimescales.jl calculates the ACF as the inverse Fourier transform of squared magnitude of Fourier transform of data for increased performance. PSD calculation is done using the periodogram method with a hamming window. In the case of missing data, ACF is calculated in time domain while accounting for the missingness of the data and PSD is calculated with the Lomb-Scargle periodogram.
- Visualization:
    - IntrinsicTimescales.jl offers built-in plotting utilities for ACFs and PSDs. For Bayesian methods, plots of posterior predictive checks are available.

A diagram shows the major features of the package can be seen in Figure 1. Additionally, in order to facilitate usage in neuroscientific community, we are working on a Python frontend which will soon be accessible in [https://github.com/duodenum96/INTpy](https://github.com/duodenum96/INTpy).

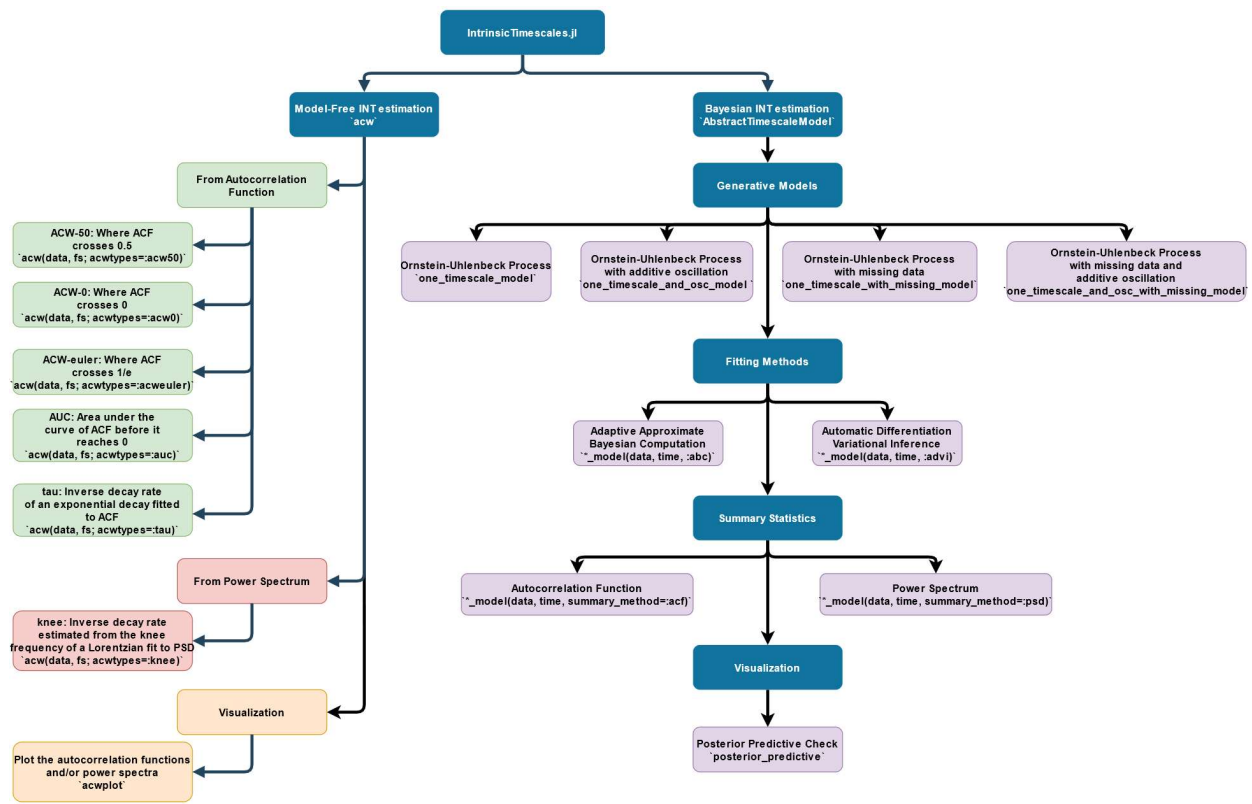

Figure 1: A diagram showing the features of the package. \* in \*_model denotes one of the models Generative Models.

## Documentation

IntrinsicTimescales.jl provides comprehensive documentation that includes detailed API references and practical tutorials. The documentation is structured in three main sections: 1) Practice tutorials that build up understanding from basic concepts to advanced methods, 2) Implementation details for both model-free and Bayesian methods, and 3) Complete API reference. Each estimation method is thoroughly documented with mathematical formulations, example code and explanations of parameters. The documentation includes code examples demonstrating proper usage of the package's features, from basic timescale estimation to advanced Bayesian inference techniques. All documentation is hosted online and integrated with Julia's built-in help system.

IntrinsicTimescales.jl aims to become a standard tool in neuroscience research by providing robust, efficient, and well-documented methods for estimating intrinsic neural timescales across different experimental paradigms and data conditions.

## Acknowledgements

There are no acknowledgments to declare and no conflicts of interest to disclose.